\documentclass[12pt]{article}

\usepackage[english]{babel}
\usepackage{graphicx,rotating}
\usepackage{fleqn,a4p,here}
\usepackage{cite,mcite,epsfig}
\usepackage{lep2rep}

\newcommand{\Qfbhad}{\ensuremath{{Q^{\mathrm{had}}_{\mathrm{FB}} }}}

\def\gappeq{\mathrel{\rlap {\raise.5ex\hbox{$>$}} {\lower.5ex\hbox{$\sim$}}}}
\def\lappeq{\mathrel{\rlap{\raise.5ex\hbox{$<$}} {\lower.5ex\hbox{$\sim$}}}}

\begin{document}

\begin{center}
\Large {EUROPEAN ORGANIZATION FOR NUCLEAR RESEARCH}
\end{center}

\begin{flushright}
       CERN-PH-TH/2004-067 \\
       UCD-EXPH/040401\\
       hep-ph/0404165 \\
       {\bf 20 April 2004}
\end{flushright}

\vskip 1cm 

\begin{center}
\Huge {\bf Precision Electroweak Tests of \\the Standard Model\\}

\vskip 1cm

\Large {Guido Altarelli\\ CERN PH-TH, Geneva 23, Switzerland}\\[10mm]

\Large {Martin~W.~Gr\"unewald\\ 
        Department of Experimental Physics\\
        University College Dublin, Dublin 4, Ireland}\\

\end{center}

\vskip 1cm

\begin{center}

{\bf Abstract}

\vskip 0.5cm

\end{center}

{ The study of electron-positron collisions at LEP, together with
  additional measurements from other experiments, in particular those
  at SLC and at the Tevatron, has allowed for tests of the electroweak
  Standard Model with unprecedented accuracy.  We review the results
  of the electroweak precision tests and their implications on the
  determination of the Standard Model parameters, in particular of the
  Higgs boson mass, and comment on the constraints for possible new
  physics effects.  }

\vfill

\begin{center}
{\em To appear in a special issue of Physics Reports dedicated
to CERN\\ on the occasion of the laboratory's 50th anniversary}
\end{center}

\vfill


\clearpage

\section{Introduction}

The experimental study of the electroweak interaction and the Standard
Model (SM) has made a quantum leap in the last 15 years. With the
advent of electron-positron colliders reaching for the first time
centre-of-mass energies of $91~\GeV$, on-shell production of the Z
boson, $\ee\to\Zzero$, allowed precision studies of Z boson properties
and the neutral weak current of electroweak interactions.  In 1989,
two $\ee$ colliders commenced operations on far away sides of the
world: the Stanford Linear Collider (SLC) at SLAC, California, USA,
and the circular Large Electron Positron collider (LEP) at CERN,
Geneva, Switzerland.  While SLC delivered collisions with a
longitudinally polarised electron beam, LEP's high luminosity made it
a true Z factory.

Five large-scale detectors collected data on $\ee$ collision
processes: SLD at SLC, and ALEPH, DELPHI, L3 and OPAL at LEP.  These
modern detectors have a typical size of 10m by 10m by 10m, surrounding
the interaction region. The detectors' high granularity and near
complete hermeticity ensure that all parts of collision events are
well measured. Dedicated luminosity monitors using Bhabha scattering
at low polar angles measured the luminosity with sub per-mille
precision, paving the way for highly precise cross section
determinations.  Owing to the superior energy and spatial resolution
of the five detectors, greatly improved by the subsequent installation
of silicon micro-vertex detectors, measurements of observables
pertaining to the electroweak interaction have been performed with
per-mille precision~\cite{LEPEWWG:2003}, unprecedented in high energy
particle physics outside QED.

This article presents the main results of the programme in electroweak
physics at SLC and LEP, covering the measurements at the Z pole but
also the second phase of LEP, 1996-2000, where W boson properties were
determined based on on-shell W-pair production, $\ee\to\WW$.  We put
the measurements by SLD and the four LEP experiments together with
relevant measurements performed at other colliders; most notably the
results from the experiments CDF and D\O, which are taking data at the
proton-antiproton collider Tevatron, on the mass of W boson and top
quark~\cite{TEVEWWG-W,TEVEWWG-top}.

\section{The Z Boson}

\begin{figure}[htbp]
\begin{center}
\includegraphics[width=0.9\textwidth]{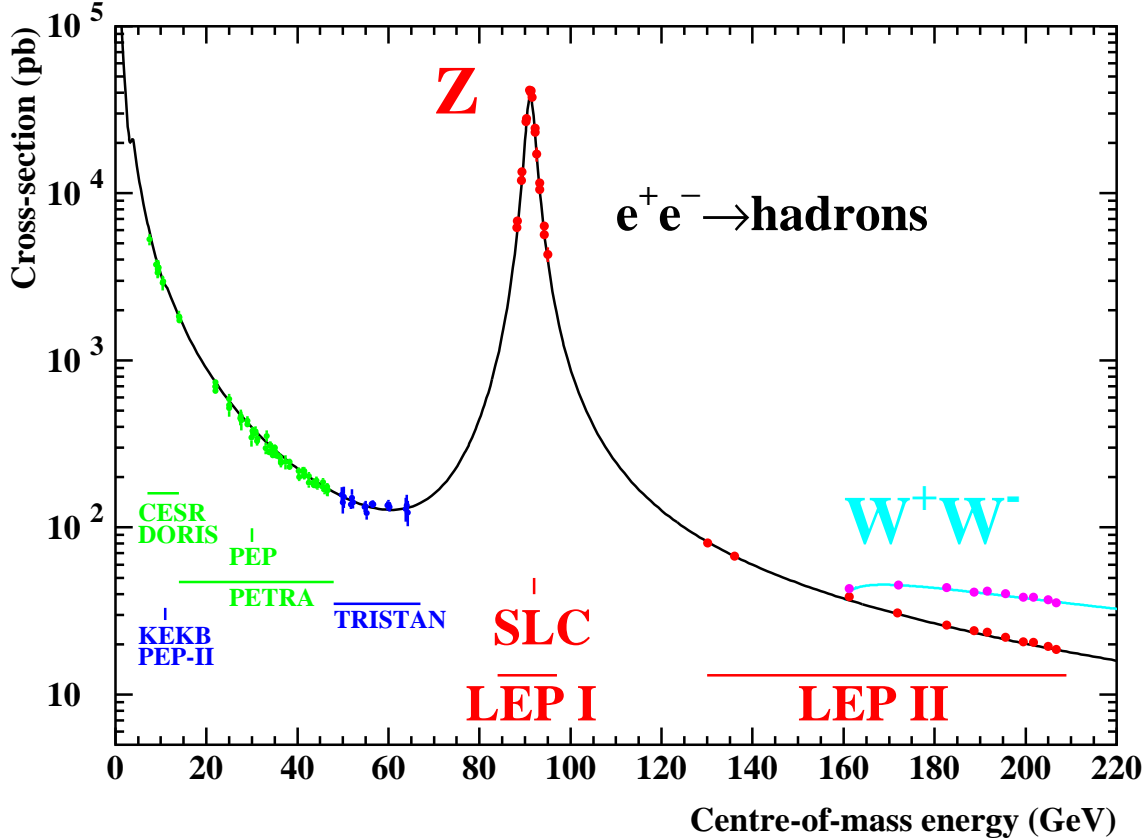}
\caption[]{The cross-section for the production of hadrons in $\ee$
annihilations. The measurements are shown as dots with error bars.
The solid line shows the prediction of the SM.}
\label{fig:intro_xhad}
\end{center}
\end{figure}

The process of electron-positron annihilation into fermion-antifermion
pairs proceeds via virtual photon and Z boson exchange.  As shown in
Figure~\ref{fig:intro_xhad}, the cross section is dominated by the
resonant formation of the Z boson at centre-of-mass energies close to
the mass of the Z boson.  While SLC mostly studied collisions at the
peak energy to maximize event yield, LEP scanned the centre-of-mass
energy region from $88~\GeV$ to $94~\GeV$.  A total of 15.5 million
hadronic events and 1.7 million lepton-pair events have been recorded
by the four LEP experiments, while SLD collected 0.6 million events
with longitudinal polarisation of the electron beam in excess of 70\%.
The three charged lepton species are analysed separately, while the
five kinematically accessible quark flavours are treated inclusively
in the hadronic final state. Special tagging methods exploiting
heavy-quark properties allow the separation of samples highly enriched
in Z decays to $\mathrm{b \bar b}$ and $\mathrm{c \bar c}$ pairs, and
thus the determination of partial decay widths and asymmetries for the
corresponding heavy-quark flavours.

Analysing the resonant Z lineshape in the various Z decay modes leads
to the determination of mass, total and partial decay widths of the Z
boson as parametrised by a relativistic Breit-Wigner with an $s$
dependent total width, $\MZ$, $\GZ$ and $\Gff$.  Owing to the precise
determination of the LEP beam energy, mass and total width of the Z
resonance are now known at the $\MeV$ level; the combination of all
results yields:
\begin{eqnarray}
\MZ & = &          9 1.1875 \pm 0.0021 ~\GeV\\
\GZ & = & \phantom{9}2.4952 \pm 0.0023 ~\GeV\,.
\end{eqnarray}
Note that the relative accuracy of $\MZ$ is in the same order as that
of the Fermi constant $\GF$.  The total width $\GZ$ corresponds to a
lifetime $\tau_Z=(2.6379\pm0.0024)10^{-25}s$.

An important aspect of the Z lineshape analysis is the determination
of the number of light neutrino flavours coupling to the Z boson. The
result is:
\begin{eqnarray}
\Nnu & = & 2.9841\pm0.0083\,,
\end{eqnarray}
about 1.9 standard deviations less than 3.  This result shows that
there are just the known three flavours; hence there exist only the
three known sequential generations of fermions (with light neutrinos),
a result with important consequences in astrophysics and cosmology.

\begin{table}[tp]
\begin{center}
  \renewcommand{\arraystretch}{1.30}
\begin{tabular}{|ll||r||r|}
\hline
&Observable& {Measurement}  & {SM fit}  \\
\hline
\hline
&$\MZ$ [\GeV{}] & $91.1875\pm0.0021\pz$ &91.1873$\pz$ \\
&$\GZ$ [\GeV{}] & $2.4952 \pm0.0023\pz$ & 2.4965$\pz$ \\
&$\shad$ [nb]   & $41.540 \pm0.037\pzz$ &41.481$\pzz$ \\
&$\Rl$          & $20.767 \pm0.025\pzz$ &20.739$\pzz$ \\
&$\Afbzl$       & $0.0171 \pm0.0010\pz$ & 0.0164$\pz$ \\
\hline
&$\cAl$~(SLD)   & $0.1513 \pm0.0021\pz$ & 0.1480$\pz$ \\
\hline
&$\cAl~(P_\tau)$& $0.1465 \pm0.0033\pz$ & 0.1480$\pz$ \\
\hline
&$\Rbz{}$       & $0.21644\pm0.00065$   & 0.21566     \\
&$\Rcz{}$       & $0.1718\pm0.0031\pz$  & 0.1723$\pz$ \\
&$\Afbzb{}$     & $0.0995\pm0.0017\pz$  & 0.1037$\pz$ \\
&$\Afbzc{}$     & $0.0713\pm0.0036\pz$  & 0.0742$\pz$ \\
&$\cAb$         & $0.922\pm 0.020\pzz$  & 0.935$\pzz$ \\
&$\cAc$         & $0.670\pm 0.026\pzz$  & 0.668$\pzz$ \\
\hline
&$\swsqeffl$
  ($\Qfbhad$)   & $0.2324\pm0.0012\pz$  & 0.23140     \\
\hline
\hline
&$\MW$ [\GeV{}]
                & $80.425\pm0.034\pzz$  &80.398$\pzz$ \\
&$\GW$ [\GeV{}]
                & $ 2.133\pm0.069\pzz$  & 2.094$\pzz$ \\
\hline
&$\Mt$ [\GeV{}] ($\pp$~\cite{TEVEWWG-top})
                & $178.0\pm4.3\pzz\pzz$ &178.1$\pzz\pzz$ \\
\hline
& $\dalhad$~\cite{bib-BP01}
                & $0.02761\pm0.00036$   & 0.02768     \\
\hline
\end{tabular}\end{center}
\caption[Overview of results]{ Summary of electroweak precision
measurements at high $Q^2$~\cite{LEPEWWG:2003}. The first block shows
the Z-pole measurements.  The second block shows additional results
from other experiments: the mass and the width of the W boson measured
at the Tevatron and at LEP-2, the mass of the top quark measured at
the Tevatron, and the the contribution to $\alqed$ of the hadronic
vacuum polarisation.  For the correlations between the measurements,
taken into account in our analysis, see~\cite{LEPEWWG:2003}. The SM
fit results are derived from the SM analysis of these 18 results, also
including constants such as the Fermi constant $\GF$ (fit 3 of
Table~\ref{tab:fit:result}), using the programs TOPAZ0~\cite{TOPAZ0}
and ZFITTER~\cite{ZFITTER}. }
\label{tab:msm:input}
\end{table}

All electroweak Z pole measurements, combining the results of the 5
experiments, are summarised in Table~\ref{tab:msm:input}.  The cross
section scale is given by the pole cross sections for the various
final states $\sigma^0$; ratios thereof correspond to ratios of
partial decay widths:
\begin{eqnarray}
\shad & = & \frac{12\pi}{\MZ^2}\ \frac{\Gee\Ghad}{\GZ^2}\,, \qquad
\Rl ~ = ~ \frac{\shad}{\slept} ~ = ~ \frac{\Ghad}{\Gll}\,, \qquad
\Rq ~ = ~ \frac{\Gqq}{\Ghad} \,.
\end{eqnarray}
Here $\Gll$ is the partial decay width for a pair of massless charged
leptons.  The partial decay width for a given fermion species contains
information about the effective vector and axial-vector coupling
constants of the neutral weak current:
\begin{eqnarray}
\Gff & = & N_C^f \frac{\GF\MZ^3}{6\sqrt{2}\pi} 
\left( \gaf^2 C_{\mathrm{Af}} + \gvf^2 C_{\mathrm{Vf}} \right) 
          +  \Delta_{\rm ew/QCD}\,,
\end{eqnarray}
where $N_C^f$ is the QCD colour factor, $C_{\mathrm{\{A,V\}f}}$ are
final-state QCD/QED correction factors also absorbing imaginary
contributions to the effective coupling constants, $\gaf$ and $\gvf$
are the real parts of the effective couplings, and $\Delta$ contains
non-factorisable mixed corrections.

\begin{figure}[htb]
\begin{center}
\includegraphics[width=0.495\textwidth]{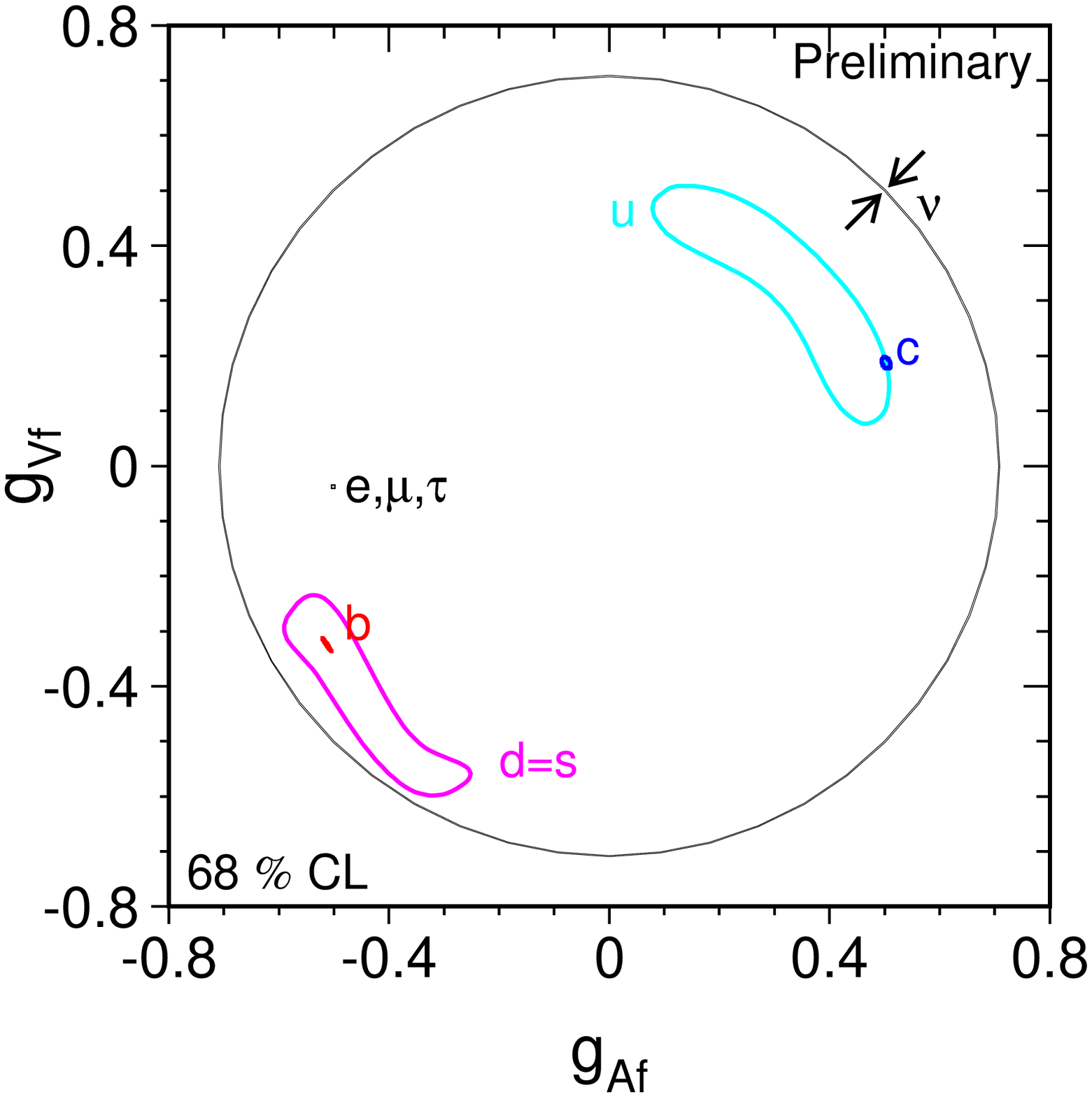} 
\hfill
\includegraphics[width=0.495\textwidth]{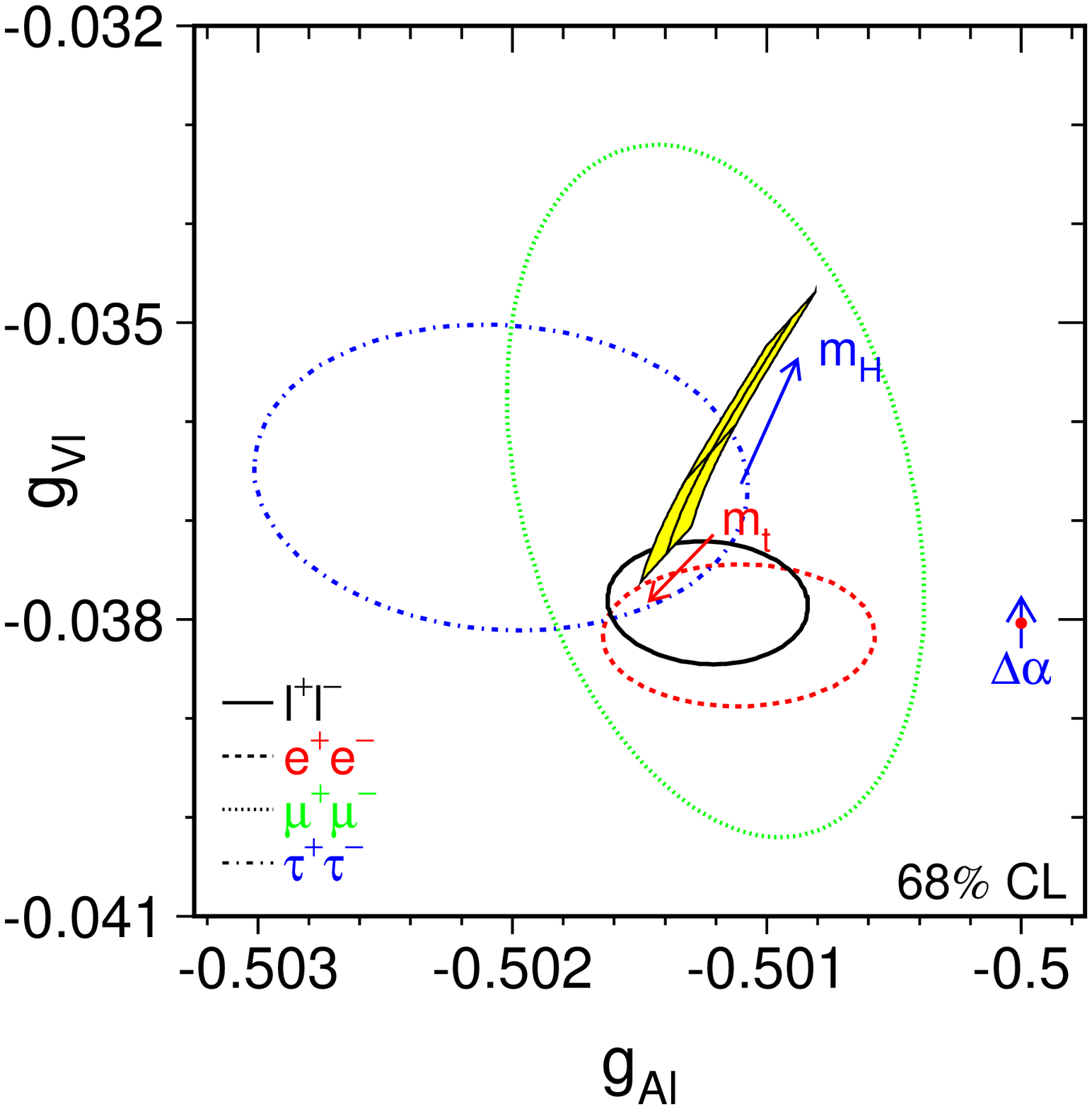}
\caption[Effective coupling constants] { Left: Effective vector and
axial-vector coupling constants for fermions. For light quarks,
identical couplings for d and s quarks are assumed in the
analysis. The allowed area for neutrinos, assuming three generations
of neutrinos with identical vector and axial-vector couplings, is a
thin ring bounded by two virtually identical circles centred at the
origin. On the scale of the left plot, the SM expectation of up and
down type quarks lie on top of the b and c allowed regions.  Right:
Effective vector and axial-vector coupling constants for leptons. The
shaded region in the lepton plot shows the predictions within the SM
for $\Mt=178.0\pm4.3~\GeV$ and $\MH=300^{+700}_{-186}~\GeV$; varying
the hadronic vacuum polarisation by $\dalhad=0.02761\pm0.00036$ yields
an additional uncertainty on the SM prediction shown by the arrow
labelled $\Delta\alpha$.  }
\label{fig:coup:gf} 
\end{center}
\end{figure}

Besides total cross sections, various types of asymmetries have been
measured.  The results of all asymmetry measurements are quoted in
terms of the asymmetry parameter $\cAf$, defined in terms of the real
parts of the effective coupling constants, $\gvf$ and $\gaf$, as:
\begin{eqnarray}
\cAf & = & 2\frac{\gvf\gaf}{\gvf^2+\gaf^2} ~ = ~ 
           2\frac{\gvf/\gaf}{1+(\gvf/\gaf)^2}\,, \qquad
\Afbzf ~ = ~ \frac{3}{4}\cAe\cAf\,.
\end{eqnarray}
The measurements are: the forward-backward asymmetry ($\Afbzf =
(3/4)\cAe\cAf$), the tau polarisation ($\cAt$) and its forward
backward asymmetry ($\cAe$) measured at LEP, as well as the left-right
and left-right forward-backward asymmetry measured at SLC ($\cAe$ and
$\cAf$, respectively).  Hence the set of partial width and asymmetry
results allows the extraction of the effective coupling constants. An
overview comparing all fermion species in the $(\gaf,\gvf)$ plane is
given in Figure~\ref{fig:coup:gf} left, while an expanded view of the
leptonic couplings is given in Figure~\ref{fig:coup:gf}
right. Compared to the situation in 1987, the accuracy of the
effective coupling constants has improved by more than a factor of
100. Lepton universality of the neutral weak current is now
established at the per-mille level.

Using the effective electroweak mixing angle, $\swsqefff$, and the
$\rho$ parameter, the effective coupling constants are given by:
\begin{eqnarray}
\gaf              & = & \sqrt{\rho}~T^f_3\,, 
\qquad 
\frac{\gvf}{\gaf} ~ = ~ 1-4|q_f|\swsqefff\,,
\end{eqnarray}
where $T^f_3$ is the third component of the weak iso-spin and $q_f$
the electric charge of the fermion. The effective electroweak mixing
angle is thus given independently of the $\rho$ parameter by the ratio
$\gvf/\gaf$ and hence in a one-to-one relation by each asymmetry
result.

\begin{figure}[htb]
\begin{center}
\includegraphics[width=0.8\textwidth]{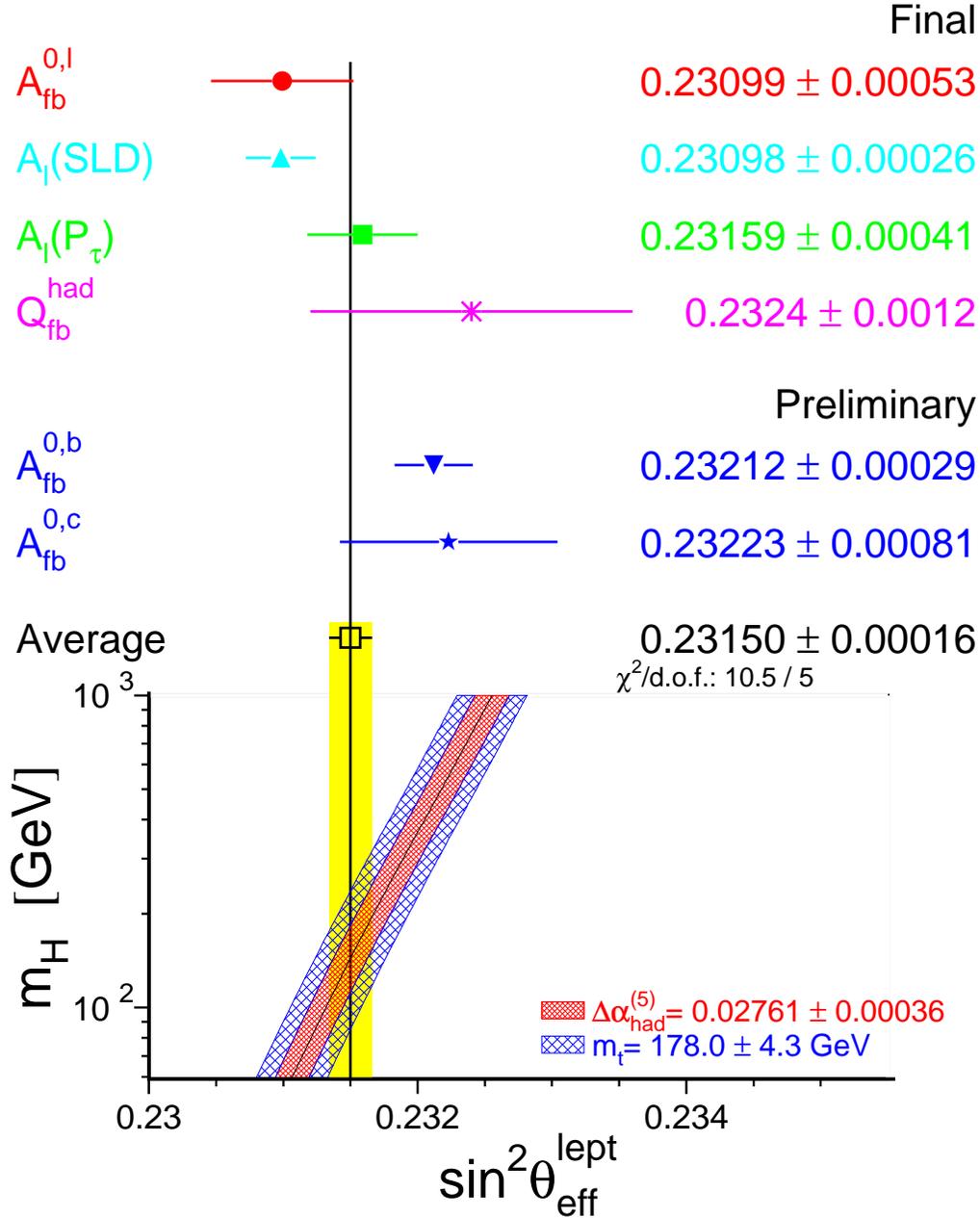}
\caption[Effective electroweak mixing angle] { Effective electroweak
mixing angle $\swsqeffl$ derived from measurement results depending on
lepton couplings only (top) and also quark couplings (bottom).  Also
shown is the prediction of $\swsqeffl$ in the SM as a function of
$\MH$, including its parametric uncertainty dominated by the
uncertainties in $\dalhad$ and $\Mt$, shown as the bands. }
\label{fig:sin2teff} 
\end{center}
\end{figure}

The various asymmetries determine the effective electroweak mixing
angle for leptons with highest sensitivity.  The results on
$\swsqeffl$ are compared in Figure~\ref{fig:sin2teff}. The weighted
average of these six results, including small correlations, is:
\begin{eqnarray}
\swsqeffl & = & 0.23150\pm0.00016 \,.
\label{eq:sin2teff}
\end{eqnarray}
Note, however, that this average has a $\chi^2$ of 10.5 for 5 degrees
of freedom, corresponding to a probability of 6.2\%. The $\chi^2$ is
pushed up by the two most precise measurements of $\swsqeffl$, namely
those derived from the measurements of $\cAl$ by SLD, dominated by the
left-right asymmetry $\ALRz$, and of the forward-backward asymmetry
measured in $\bb$ production at LEP, $\Afbzb$, which differ by about
2.9 standard deviations. No experimental effect in either measurement
has been identified to explain this, thus the difference is presumably
either a statistical fluctuation or a hint for new physics, further
discussed below.

\section{The W Boson}

With the installation of superconducting RF cavities, the
centre-of-mass energy for $\ee$ collisions provided by LEP was more
than doubled. From 1996 to the end of LEP running in the year 2000,
the centre-of-mass energy increased from $160~\GeV$, the kinematic
threshold of W-pair production, up to $209~\GeV$. More than 40,000
W-pair events in all W decay modes, $\WWtoqqqq$, $\WWtoqqlv$ and
$\WW\to\lv\lv$, have been recorded by the four LEP experiments. Among
the many measurements of W boson properties, the W-pair production
cross section and the mass and total width of the W boson are of
central importance to the electroweak SM.

\begin{figure}[t]
\begin{center}
\includegraphics[width=0.8\textwidth]{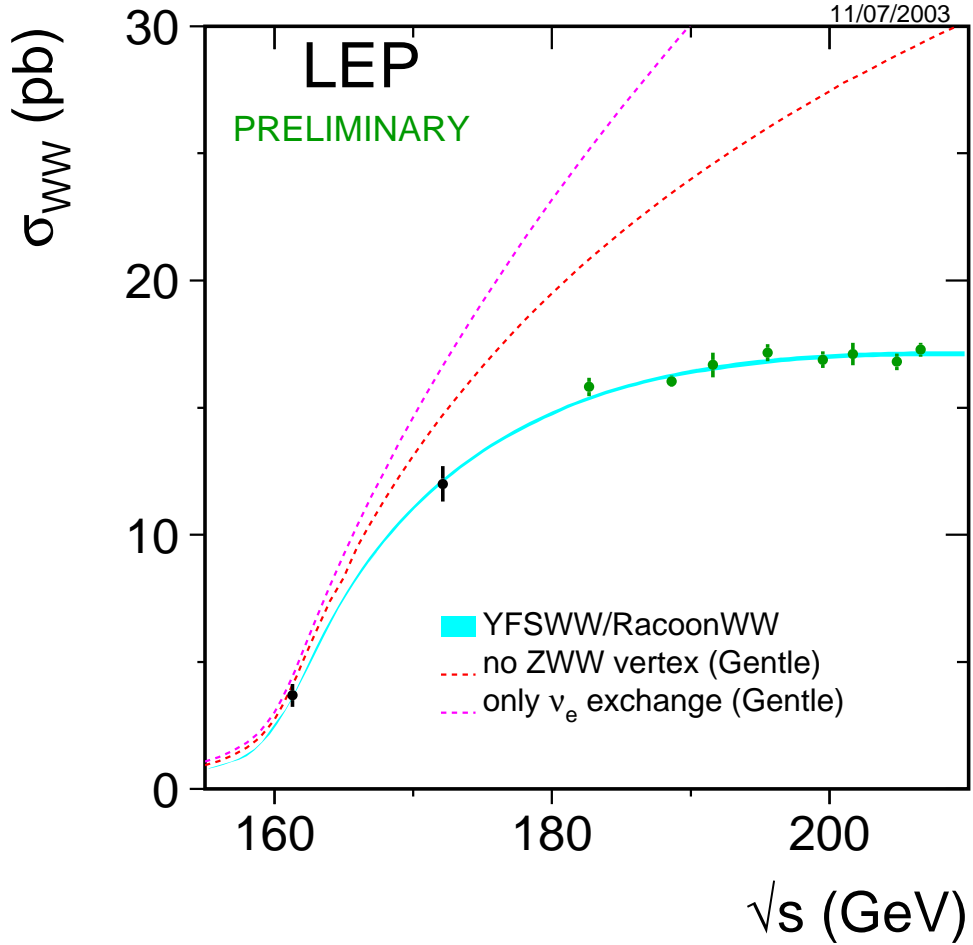}
\caption[W-pair production cross section] { The measured W-pair
production cross section compared to the SM and alternative theories
not including trilinear gauge couplings.}
\label{fig:WWxsec} 
\end{center}
\end{figure}

The cross section for W-pair production is shown in
Figure~\ref{fig:WWxsec}~\cite{LEPEWWG:2003}. Trilinear gauge couplings
between the electroweak gauge bosons $\gamma$, W and Z, as prediected
by the electroweak SM, are required to explain the cross sections
measured as a function of $\sqrt{s}$.  The mass and width of the W
boson is mesured by reconstructing the invariant mass of its decay
products.  Monte Carlo events generated with a known W-boson mass
distribution are reweighted in order to obtain the best fit to the
distribution observed in data, yielding a measurement of $\MW$ and
$\GW$.  The events of the type $\WWtoqqlv$ dominate the mass
determination.  The channel $\WWtoqqqq$ is less precise due to
potentially large final-state interconnection effects arising from
cross talk between the two hadronic systems of the decaying W
bosons. Effects such as colour reconnection, or Bose-Einstein
correlation effects between final-state hadrons, may spoil the
identification of the invariant mass of the decay products with the
invariant mass of the decaying W bosons.

Combining all LEP-2 results, most still preliminary, the best values
are~\cite{LEPEWWG:2003}:
\begin{eqnarray}
\MW & = &          8 0.412  \pm 0.042  ~\GeV\\
\GW & = & \phantom{8}2.150  \pm 0.091  ~\GeV\,,
\end{eqnarray}
in very good agreement with the results from the CDF and D\O\
experiments at the Tevatron collider~\cite{TEVEWWG-W}.  The combined
LEP-2 and Tevatron results are reported in Table~\ref{tab:msm:input}.

\section{Interpretation within the Standard Model}

For the analysis of electroweak data in the SM one starts from the
input parameters: as in any renormalisable theory masses and couplings
have to be specified from outside. One can trade one parameter for
another and this freedom is used to select the best measured ones as
input parameters. As a result, some of them, $\alpha$, $G_F$ and
$\MZ$, are very precisely known, some other ones, $m_{f_{light}}$,
$\Mt$ and $\alpha_s(m_Z)$ are far less well determined while $\MH$ is
largely unknown.  Note that the new combined CDF and D\O\ value for
$\Mt$~\cite{TEVEWWG-top}, as listed in Table~\ref{tab:msm:input}, is
higher than the previous average by nearly one standard deviation.

Among the light fermions, the quark masses are badly known, but
fortunately, for the calculation of radiative corrections, they can be
replaced by $\alpha(m_Z)$, the value of the QED running coupling at
the Z mass scale. The value of the hadronic contribution to the
running, $\dalhad$, reported in Table~\ref{tab:msm:input}, is obtained
through dispersion relations from the data on $\ee\to\rm{hadrons}$ at
low centre-of-mass energies~\cite{bib-BP01}. From the input parameters
one computes the radiative corrections to a sufficient precision to
match the experimental accuracy. Then one compares the theoretical
predictions and the data for the numerous observables which have been
measured, checks the consistency of the theory and derives constraints
on $\Mt$, $\alfmz$ and $\MH$.

The computed radiative corrections include the complete set of
one-loop diagrams, plus some selected large subsets of two-loop
diagrams and some sequences of resummed large terms of all orders
(large logarithms and Dyson resummations). In particular large
logarithms, e.g., terms of the form $(\alpha/\pi ~{\rm
ln}~(m_Z/m_{f_\ell}))^n$ where $f_{\ell}$ is a light fermion, are
resummed by well-known and consolidated techniques based on the
renormalisation group. For example, large logarithms dominate the
running of $\alpha$ from $m_e$, the electron mass, up to $\MZ$, which
is a $6\%$ effect, much larger than the few per-mille contributions of
purely weak loops.  Also, large logs from initial state radiation
dramatically distort the line shape of the Z resonance observed at
LEP-1 and SLC and must be accurately taken into account in the
measurement of the Z mass and total width.

Among the one loop EW radiative corrections a remarkable class of
contributions are those terms that increase quadratically with the top
mass.  The large sensitivity of radiative corrections to $m_t$ arises
from the existence of these terms. The quadratic dependence on $m_t$
(and possibly on other widely broken isospin multiplets from new
physics) arises because, in spontaneously broken gauge theories, heavy
loops do not decouple. On the contrary, in QED or QCD, the running of
$\alpha$ and $\alpha_s$ at a scale $Q$ is not affected by heavy quarks
with mass $M \gg Q$. According to an intuitive decoupling
theorem~\cite{AppCar}, diagrams with heavy virtual particles of mass
$M$ can be ignored for $Q \ll M$ provided that the couplings do not
grow with $M$ and that the theory with no heavy particles is still
renormalizable. In the spontaneously broken EW gauge theories both
requirements are violated. First, one important difference with
respect to unbroken gauge theories is in the longitudinal modes of
weak gauge bosons. These modes are generated by the Higgs mechanism,
and their couplings grow with masses (as is also the case for the
physical Higgs couplings). Second, the theory without the top quark is
no more renormalisable because the gauge symmetry is broken if the b
quark is left with no partner (while its couplings show that the weak
isospin is 1/2). Because of non decoupling precision tests of the
electroweak theory may be sensitive to new physics even if the new
particles are too heavy for their direct production.

While radiative corrections are quite sensitive to the top mass, they
are unfortunately much less dependent on the Higgs mass. If they were
sufficiently sensitive, by now we would precisely know the mass of the
Higgs. However, the dependence of one loop diagrams on $\MH$ is only
logarithmic: $\sim \GF\MW^2\log(\MH^2/\MW^2)$. Quadratic terms $\sim
\GF^2\MH^2$ only appear at two loops and are too small to be
important. The difference with the top case is that $\Mt^2-\Mb^2$ is a
direct breaking of the gauge symmetry that already affects the
relevant one loop diagrams, while the Higgs couplings to gauge bosons
are "custodial-SU(2)" symmetric in lowest order.

We now discuss fitting the data in the SM. One can think of different
types of fit, depending on which experimental results are included or
which answers one wants to obtain. For example, in
Table~\ref{tab:fit:result} we present in column~1 a fit of all Z pole
data plus $\MW$ and $\GW$ (this is interesting as it shows the value
of $m_t$ obtained indirectly from radiative corrections, to be
compared with the value of $m_t$ measured in production experiments),
in column~2 a fit of all Z pole data plus $\Mt$ (here it is $\MW$
which is indirectly determined), and, finally, in column~3 a fit of
all the data listed in Table~\ref{tab:msm:input} (which is the most
relevant fit for constraining $\MH$).  From the fit in column~1 of
Table~\ref{tab:fit:result} we see that the extracted value of $\Mt$ is
in perfect agreement with the direct measurement (see
Table~\ref{tab:msm:input}).  Similarly we see that the experimental
measurement of $\MW$ in Table~\ref{tab:msm:input} is larger by about
one standard deviation with respect to the value from the fit in
column~2.  We have seen that quantum corrections depend only
logarithmically on $\MH$.  In spite of this small sensitivity, the
measurements are precise enough that one still obtains a quantitative
indication of the mass range. From the fit in column~3 we obtain:
$\log_{10}{\MH(\GeV)}=2.05\pm 0.20$ (or $\MH=113^{+62}_{-42}~\GeV$).
This result on the Higgs mass is particularly remarkable. The value of
$\log_{10}{\MH(\GeV)}$ is right on top of the small window between
$\sim 2$ and $\sim 3$ which is allowed, on the one side, by the direct
search limit ($\MH\gappeq 114~\GeV$ from LEP-2~\cite{LEP2:MH-LIMIT}),
and, on the other side, by the theoretical upper limit on the Higgs
mass in the minimal SM, $\MH\lappeq 600-800~\GeV$ \cite{HR}.

\begin{table}[tb]
\begin{center}
\renewcommand{\arraystretch}{1.3}
\begin{tabular}{|l||c|c|c|}
\hline 
Fit       & 1 & 2 & 3 \\
\hline
\hline
Measurements      &$\MW$,~$\GW$          &$\Mt$            &$\Mt,~\MW,~\GW$\\
\hline
\hline
$\Mt~(\GeV)$      &$178.5^{+11.0}_{-8.5}$&$177.2\pm4.1$    &$178.1\pm3.9$\\
$\MH~(\GeV)$      &$117^{+162}_{-62}$    &$129^{+76}_{-50}$&$113^{+62}_{-42}$\\
$\log~[\MH(\GeV)]$&$2.07^{+0.38}_{-0.33}$&$2.11\pm0.21$    &$2.05\pm0.20$ \\
$\alpha_s(\MZ)$   &$0.1187\pm0.0027$     &$0.1190\pm0.0027$&$0.1186\pm0.0027$\\
\hline
$\chi^2/dof$      &$16.3/12$             &$15.0/11$        &$16.3/13$ \\
\hline
$\MW~(\MeV)$      &                      &$80386 \pm 23$   & \\
\hline
\end{tabular}
\caption[]{ Standard Model fits of electroweak data. All fits use the
Z pole results and $\dalhad$ as listed in Table~\ref{tab:msm:input},
also including constants such as the Fermi constant $\GF$. In
addition, the measurements listed in each column are included as
well. For fit~2, the expected W mass is also shown. For details on the
fit procedure, using the programs TOPAZ0~\cite{TOPAZ0} and
ZFITTER~\cite{ZFITTER}, see~\cite{LEPEWWG:2003}.}
\label{tab:fit:result}
\end{center}
\end{table} 

\begin{table}[tb]
\begin{center}
  \renewcommand{\arraystretch}{1.30}
\begin{tabular}{|ll||r||r|}
\hline
&Observable& {Measurement}  & {SM fit}  \\
\hline
\hline
&$\swsq$        ($\nu$N~\cite{bib-NuTeV-final})
                & $0.2277\pm0.0016\pz$  & 0.2226$\pz$ \\
\hline
&$\QWCs$ (APV~\cite{QWCs:theo:2003})
                & $-72.84\pm0.49\pzz\pz$& $-72.91\pzz\pz$ \\
\hline
&$\swsqeffl$   ($\mathrm{e^-e^-}$~\cite{E158})
                & $0.2296\pm0.0023\pz$  & 0.2314$\pz$ \\
\hline
\end{tabular}\end{center}
\caption[Overview of results]{ Summary of other electroweak precision
measurements, namely the measurements of the on-shell electroweak
mixing angle in neutrino-nucleon scattering, the weak charge of cesium
measured in an atomic parity violation experiment, and the effective
weak mixing angle measured in Moller scattering, all performed in
processes at low $Q^2$.  The SM predictions are derived from fit 3 of
Table~\ref{tab:fit:result}.  Good agreement of the prediction with the
measurement is found except for $\nu$N. }
\label{tab:msm:prediction}
\end{table}

Thus the whole picture of a perturbative theory with a fundamental
Higgs is well supported by the data on radiative corrections. It is
important that there is a clear indication for a particularly light
Higgs: at $95\%$ c.l. $m_H\lappeq 237~\GeV$.  This is quite
encouraging for the ongoing search for the Higgs particle.  More
general, if the Higgs couplings are removed from the Lagrangian the
resulting theory is non renormalisable. A cutoff $\Lambda$ must be
introduced. In the quantum corrections $\log{m_H}$ is then replaced by
$\log{\Lambda}$ plus a constant. The precise determination of the
associated finite terms would be lost (that is, the value of the mass
in the denominator in the argument of the logarithm).  A heavy Higgs
would need some unfortunate conspiracy: the finite terms, different in
the new theory from those of the SM, should accidentally compensate
for the heavy Higgs in a few key parameters of the radiative
corrections (mainly $\epsilon_1$ and $\epsilon_3$, see, for example,
\cite{eps}).  Alternatively, additional new physics, for example in
the form of effective contact terms added to the minimal SM
lagrangian, should accidentally do the compensation, which again needs
some sort of conspiracy.

In Table~\ref{tab:msm:prediction} we collect the results on low energy
precision tests of the SM obtained from neutrino and antineutrino deep
inelastic scattering (NuTeV~\cite{bib-NuTeV-final}), parity violation
in Cs atoms (APV~\cite{QWCs:theo:2003}) and the recent measurement of
the parity-violating asymmetry in Moller scattering \cite{E158}.  The
experimental results are compared with the predictions from the fit in
column~3 of Table~\ref{tab:fit:result}.  We see the agreement is good
except for the NuTeV result that shows a deviation by three standard
deviations.  The NuTeV measurement is quoted as a measurement of
$\swsq=1-\MW^2/\MZ^2$ from the ratio of neutral to charged current
deep inelastic cross-sections from $\nu_{\mu}$ and $\bar{\nu}_{\mu}$
using the Fermilab beams. There is growing evidence that the NuTeV
anomaly could simply arise from an underestimation of the theoretical
uncertainty in the QCD analysis needed to extract $\swsq$.  In fact,
the lowest order QCD parton formalism on which the analysis has been
based is too crude to match the experimental accuracy.  In particular
a small asymmetry in the momentum carried by the strange and
antistrange quarks, $s-\bar s$, could have a large effect
\cite{DFGRS}. A tiny violation of isospin symmetry in parton
distributions, too small to be seen elsewhere, can similarly be of
some importance. In conclusion we believe the discrepancy has more to
teach about the QCD parton densities than about the electroweak
theory.

When confronted with these results, on the whole the SM performs
rather well, so that it is fair to say that no clear indication for
new physics emerges from the data. However, as already mentioned, one
problem is that the two most precise measurements of $\swsqeffl$ from
$\ALR$ and $\Afbzb$ differ nearly three standard deviations.  In
general, there appears to be a discrepancy between $\swsqeffl$
measured from leptonic asymmetries ($(\sin^2\theta_{\rm eff})_l$) and
from hadronic asymmetries ($(\sin^2\theta_{\rm eff})_h$), see also
Figure~\ref{fig:sin2teff}. In fact, the result from $\ALR$ is in good
agreement with the leptonic asymmetries measured at LEP, while all
hadronic asymmetries, though their errors are large, are better
compatible with the result of $\Afbzb$.

\begin{figure}[t]
\begin{center}
\includegraphics[width=0.7\textwidth]{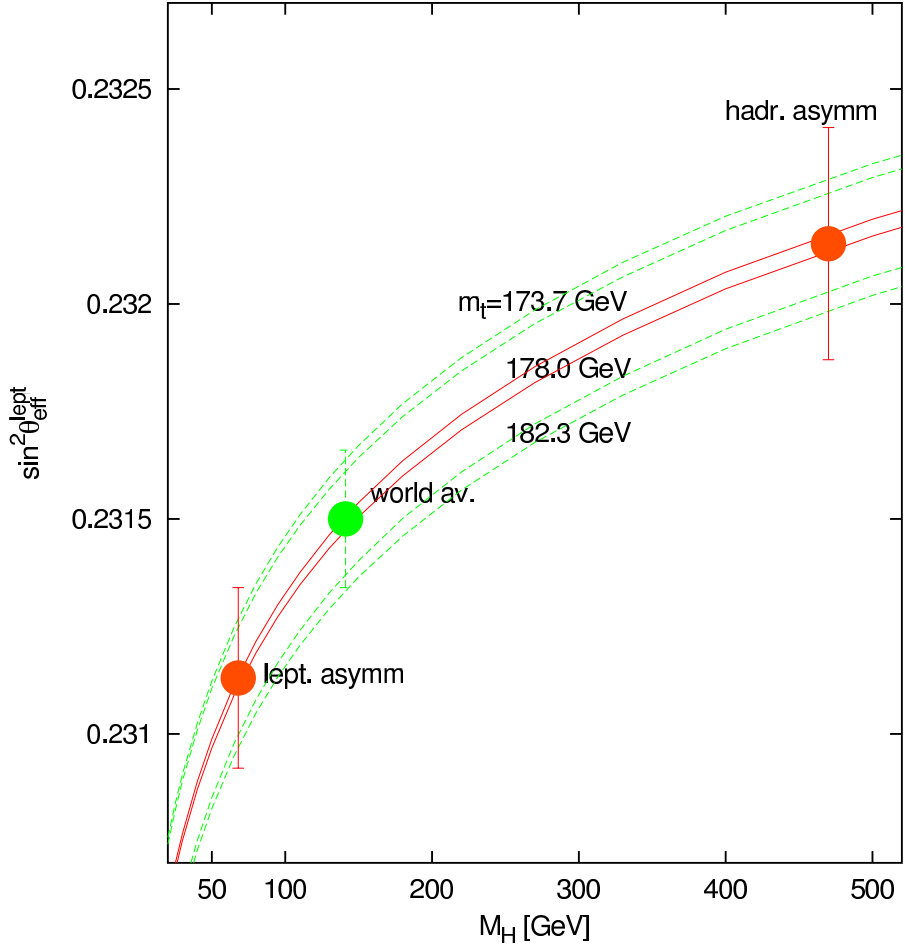}
\caption[]{ The data for $\sin^2\theta_{\rm eff}^{\rm lept}$ are
plotted vs $m_H$. For presentation purposes the measured points are
shown each at the $m_H$ value that would ideally correspond to it
given the central value of $m_t$ (updated from \cite{P-Gambino}).}
\label{fig:higgs}
\end{center}
\end{figure} 

The situation is shown in Figure~\ref{fig:higgs}~\cite{P-Gambino}.
The values of $(\sin^2\theta_{\rm eff})_l$, $(\sin^2\theta_{\rm
eff})_h$ and their formal combination are shown each at the $\MH$
value that would correspond to it given the central value of $\Mt$.
Of course, the value for $\MH$ indicated by each $\swsqeffl$ has an
horizontal ambiguity determined by the measurement error and the width
of the $\pm1\sigma$ band for $\Mt$.  Even taking this spread into
account it is clear that the implications on $\MH$ are sizably
different.  One might imagine that some new physics effect could be
hidden in the $\mathrm{Z b \bar b}$ vertex.  Like for the top quark
mass there could be other non decoupling effects from new heavy states
or a mixing of the b quark with some other heavy quark.  However, it
is well known that this discrepancy is not easily explained in terms
of some new physics effect in the $\mathrm{Z b \bar b}$ vertex. In
fact, $\Afbzb$ is the product of lepton- and b-asymmetry factors:
$\Afbzb=(3/4)\cAe\cAb$.  The sensitivity of $\Afbzb$ to $\cAb$ is
limited, because the $\cAe$ factor is small, so that a rather large
change of the b-quark couplings with respect to the SM is needed in
order to reproduce the measured discrepancy (precisely a $\sim 30\%$
change in the right-handed coupling, an effect too large to be a loop
effect but which could be produced at the tree level, e.g., by mixing
of the b quark with a new heavy vectorlike quark \cite{CTW}).  But
then this effect should normally also appear in the direct measurement
of $\cAb$ performed at SLD using the left-right polarized b asymmetry,
even within the moderate precision of this result, and it should also
be manifest in the accurate measurement of $\Rb \propto
g_{\mathrm{Rb}}^2+g_{\mathrm{Lb}}^2$.  The measurements of neither
$\cAb$ nor $\Rb$ confirm the need of a new effect. Even introducing an
ad hoc mixing the overall fit is not terribly good, but we cannot
exclude this possibility completely.  Alternatively, the observed
discrepancy could be due to a large statistical fluctuation or an
unknown experimental problem. The ambiguity in the measured value of
$\swsqeffl$ could thus be larger than the nominal error, reported in
Equation~\ref{eq:sin2teff}, obtained from averaging all the existing
determinations.

\begin{figure}[t]
\begin{center}
\includegraphics[width=0.8\textwidth]{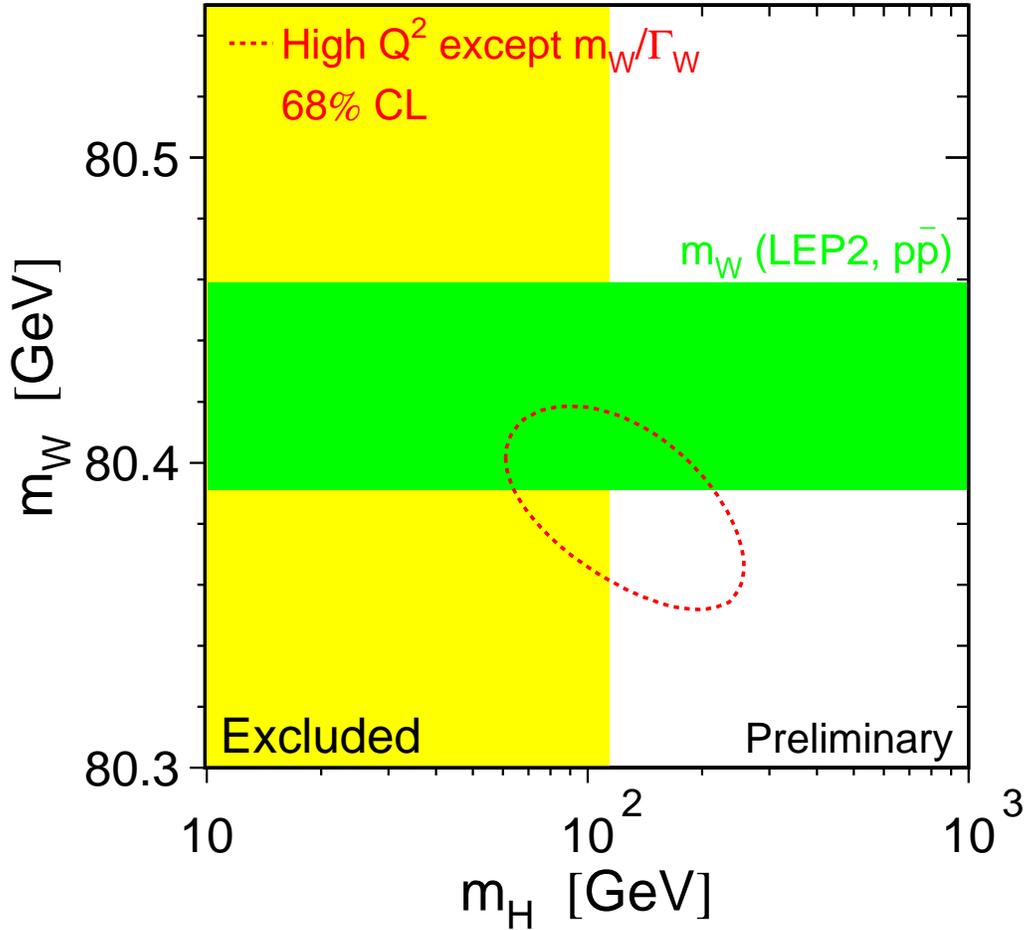}
\caption[]{Contour curve of 68\% probability in the $(\MW,\MH)$ plane
derived from fit~2 of Table~\ref{tab:fit:result}. The direct
experimental measurement, not included in the fit, is shown as the
horizontal band of width $\pm1$ standard deviation. The vertical band
shows the 95\% confidence level exclusion limit on $\MH$ of
$114~\GeV$~\cite{LEP2:MH-LIMIT}. }
\label{fig:MH-MW}
\end{center}
\end{figure}

We have already observed that the experimental value of $\MW$ (with
good agreement between LEP and the Tevatron) is a bit high compared to
the SM prediction (see Figure~\ref{fig:MH-MW}). The value of $\MH$
indicated by $\MW$ is on the low side, just in the same interval as
for $\sin^2\theta_{\rm eff}^{\rm lept}$ measured from leptonic
asymmetries.  It is interesting that the new value of $\Mt$
considerably relaxes the previous tension between the experimental
values of $\MW$ and $\sin^2\theta_{\rm eff}^{\rm lept}$ measured from
leptonic asymmetries on one side and the lower limit on $\MH$ from
direct searches on the other side~\cite{cha,ACGGR}.  This is also
apparent from Figure~\ref{fig:MH-MW}.

The main lesson of precision tests of the standard electroweak theory
can be summarised as follows. The couplings of quark and leptons to
the weak gauge bosons W$^{\pm}$ and Z are indeed precisely those
prescribed by the gauge symmetry. The accuracy of a few per-mille for
these tests implies that, not only the tree level, but also the
structure of quantum corrections has been verified. To a lesser
accuracy the triple gauge vertices $\gamma\WW$ and Z$\WW$ have also
been found in agreement with the specific prediction of the
$SU(2)\bigotimes U(1)$ gauge theory. This means that it has been
verified that the gauge symmetry is unbroken in the vertices of the
theory: the currents are indeed conserved. Yet there is obvious
evidence that the symmetry is otherwise badly broken in the
masses. Thus the currents are conserved but the spectrum of particle
states is not at all symmetric. This is a clear signal of spontaneous
symmetry breaking. The practical implementation of spontaneous
symmetry breaking in a gauge theory is via the Higgs mechanism. The
Higgs sector of the SM is still very much untested. What has been
tested is the relation $\MW^2=\MZ^2\cwsq$, modified by computable
radiative corrections. This relation means that the effective Higgs
(be it fundamental or composite) is indeed a weak isospin doublet.
The Higgs particle has not been found but in the SM its mass can well
be larger than the present direct lower limit $m_H\gappeq114~\GeV$
obtained from direct searches at LEP-2.  The radiative corrections
computed in the SM when compared to the data on precision electroweak
tests lead to a clear indication for a light Higgs, not too far from
the present lower bound. No signal of new physics has been
found. However, to make a light Higgs natural in presence of quantum
fluctuations new physics should not be too far. This is encouraging
for the LHC that should experimentally clarify the problem of the
electroweak symmetry breaking sector and search for physics beyond the
SM.

\clearpage

\bibliographystyle{lep2unsrt}
\bibliography{cern50pr}

\end{document}